\documentclass[aps,prb,preprint,12pt]{revtex4-2}
\usepackage{CJK}
\usepackage{float}
\usepackage{graphicx,amsmath,amssymb,bm}
\usepackage{color}
\usepackage{appendix}
\usepackage{ulem} 
\usepackage{amsmath}
\usepackage{natbib}
\usepackage{hyperref}
\usepackage{listings}
\usepackage{caption}
\usepackage{xcolor}

\definecolor{json_key}{RGB}{0,0,255}
\definecolor{json_value}{RGB}{0,128,0}
\definecolor{json_string}{RGB}{255,0,0}
\definecolor{bashkeyword}{RGB}{0,0,255}
\definecolor{bashstring}{RGB}{42,0.0,255}
\definecolor{bashcomment}{RGB}{0,128,0}
\definecolor{bashbackground}{RGB}{255,255,255}

\lstdefinestyle{bashstyle}{
    language=bash,
    basicstyle=\small\ttfamily,
    backgroundcolor=\color{bashbackground},
    keywordstyle=\color{bashkeyword},
    stringstyle=\color{bashstring},
    commentstyle=\color{bashcomment},
    breaklines=true,
    showstringspaces=false,
    deletekeywords={for,in},
    emph={pressure}, 
    emphstyle=\color{black}, 
    frame=tb, 
    tabsize=2, 
    showtabs=false,
    morekeywords={mkdir, cd, ls, process}, 
}
\lstdefinelanguage{json}{
    basicstyle=\small\ttfamily,
    numberstyle=\tiny,
    numbersep=8pt,
    showstringspaces=false,
    breaklines=true,
    frame=lines,
    backgroundcolor=\color{gray!10},
    literate=
     *{0}{{{\color{json_value}0}}}{1}
      {1}{{{\color{json_value}1}}}{1}
      {2}{{{\color{json_value}2}}}{1}
      {3}{{{\color{json_value}3}}}{1}
      {4}{{{\color{json_value}4}}}{1}
      {5}{{{\color{json_value}5}}}{1}
      {6}{{{\color{json_value}6}}}{1}
      {7}{{{\color{json_value}7}}}{1}
      {8}{{{\color{json_value}8}}}{1}
      {9}{{{\color{json_value}9}}}{1}
      {:}{{{\color{json_key}{:}}}}{1}
      {,}{{{\color{json_key}{,}}}}{1}
      {\{}{{{\color{json_key}{\{}}}}{1}
      {\}}{{{\color{json_key}{\}}}}}{1}
      {[}{{{\color{json_key}{[}}}}{1}
      {]}{{{\color{json_key}{]}}}}{1},
}

\usepackage{amssymb}

\newcounter{bla}

\begin{document}



\title{HTESP (High-throughput electronic structure package): a Package for high-throughput \textit{ab initio} calculations}


\author{Niraj K. Nepal$^1$}
\email[]{nnepal@ameslab.gov}
\author{Paul C. Canfield$^{1,2}$}
\author{Lin-Lin Wang$^1$}
\email[]{llw@ameslab.gov}

\affiliation{[1] Ames National Laboratory, Ames, Iowa 50011, USA}
\affiliation{[2] Department of Physics and Astronomy, Iowa State University, Ames, Iowa 50011, USA}

\begin{abstract}
High-throughput $ab$ $initio$ calculations are the indispensable parts of data-driven discovery of new materials with desirable properties, as reflected in the establishment of several online material databases. The accumulation of extensive theoretical data through computations enables data-driven discovery by constructing machine learning and artificial intelligence models to predict novel compounds and forecast their properties. Efficient usage and extraction of data from these existing online material databases can accelerate the next stage materials discovery that targets different and more advanced properties, such as electron-phonon coupling for phonon-mediated superconductivity. However, extracting data from these databases, generating tailored input files for different $ab$ $initio$ calculations, performing such calculations, and analyzing new results can be demanding tasks. Here, we introduce a software package named ``HTESP” (High-Throughput Electronic Structure Package) written in Python and Bash languages, which automates the entire workflow including data extraction, input file generation, calculation submission, result collection and plotting. Our HTESP will help speed up future computational materials discovery processes.

\end{abstract}
\maketitle

\section{Introduction}
High-throughput $ab$ $initio$ calculations of a wide range of properties for ground state and beyond, using density functional theory (DFT) and related methods, are currently receiving significant attention. This heightened interest is the result of active research in data-driven approaches aimed at discovering new materials with desirable properties. The accumulation of extensive theoretical data through computations enables data-driven discovery by constructing machine learning (ML) and artificial intelligence (AI) models to predict novel compounds and forecast their properties \cite{alphamat,matgpt}. With the introduction of several databases such as materials project (MP) \cite{JOHCRD13_MP,OCJBGCP15,Pymatgen13}, Open Quantum Materials Database (OQMD) \cite{SKAMW13,oqmd15}, and the automatic flow of materials discovery (AFLOW) \cite{aflow12,aflux17}, extracting existing data on materials and properties can accelerate the next stage materials discovery targeting different properties, such as electron-phonon coupling (EPC) for phonon-mediated superconductivity (SC). However, obtaining data from these databases, generating tailored input files for further calculations, performing such calculations, and analyzing new results can be demanding tasks. { Currently, several packages like atomate2 \cite{atomate2}, AiiDA \cite{aiida}, and VASPKIT \cite{vaspkit} exist to facilitate high-throughput frameworks, but often targeting a single database, one electronic structure code, or specific analyses. In this paper, we present a software package named “HTESP” (High-Throughput Electronic Structure Package), which is written in Python and Bash languages. With the advent of our packages, researchers will have more options to integrate various elements into a single platform. Users can enhance the package's functionality themselves by utilizing simple bash command-line scripts and corresponding Python scripts.}\\

The process of calculating phonon-mediated SC properties is highly intricate with density functional perturbation theory (DFPT) \cite{BDDG01,D01}. It encompasses a series of steps, starting from the computation of ground-state properties, phonons, EPC, and various subsequent post-processing tasks, all of which, with isotropic Eliashberg's approximation, are implemented within the Quantum Espresso (QE) code\cite{QE09,QE217}. In addition to computing SC properties such as the overall EPC strength ($\lambda$), logarithmic average of phonon frequency ($\omega_{log}$), and critical temperature of superconductivity (T$_c$), it is imperative to examine the thermodynamic phase stability (convex hull analysis) and band structures (electronic and phonon dispersion, spectral functions, and more properties) of materials. Conducting these analyses is essential to determine the viability of a material to facilitate phonon-mediated SC. Additionally, it's worth noting that these properties can be further enhanced by applying pressure, substitutional alloying, or doping for the prospective candidates. While there are some utility codes that support high-throughput computational frameworks, there is not a single package capable of performing all the high-throughput tasks mentioned above especially via command-line interface on EPC calculations. Our HTESP package has been developed to fulfill this need and more. In the package, the Python scripts are written for tasks such as generating initial inputs, extracting data, conducting  analysis, and plotting, while the Bash scripts are utilized for computational tasks, file processing during intermediate stages, and data extraction from calculations. Although the package was initially written for performing tasks on phonon-mediated SC calculations, it has various other useful functionalities. { For example, one can also compute elastic properties automatically via command-line with algorithm implemented in Pymatgen \cite{Pymatgen13, ong2010thermal, ong2008li}. Using the HTESP package, we have calculated the elastic properties of EuAl$_4$, SrAl$_4$, and BaAl$_4$, in a recently published work\cite{WNC24}. This enables us to establish a connection between the instability in the transverse acoustic phonon mode and the shearing modulus, helping to clarify the presence of charge density wave (CDW) in the first two compounds and its absence in the last one \cite{WNC24}. Similarly, one of the functionalities of the code is to conduct structural relaxation on structures obtained by distorting the structure according to the phonon eigenmodes, examining systems under hydrostatic pressure, and employing various electronic smearing techniques. In a recent study \cite{NCW24}, we leveraged this capability to uncover the low-symmetry structure of Y$_2$C$_3$ through distortion according to the imaginary phonon modes at $\Gamma$ point, followed by an investigation into its EPC properties. Additionally, besides distortion, the application of pressure and the utilization of large smearing can also stabilize phonons, facilitating the study of superconductivity, which can be automatically conducted using this code.} Here is the current list of the tasks that this package can accomplish, as a high-throughput workflow.\\

\begin{itemize}
    \item Retrieving and formatting input files from MP, AFLOW, and OQMD databases for QE and VASP\cite{vasp1,vasp2} calculations.

    \item Conducting ground-state calculations, including structure relaxation, band structure with atomic and orbital projections, and density of states (DOS) calculations, with comprehensive convergence tests.

    \item Performing EPC calculations and investigating superconductivity in the isotropic Eliashberg approximation, with spectral function ($\alpha^2$F) plotting, phonon dispersion analysis with atomic projections.

   \item Conducting phonon and thermodynamic calculations using the phonopy package\cite{phono1,phono2}.

    \item Executing ground-state calculations to construct thermodynamic phase diagrams (ground state convex hulls)\cite{ong2010thermal,ong2008li}.

   \item Performing Fermi surface calculations utilizing VASP output file ``vasprun.xml" and the IFERMI\cite{ifermi} package for visualization and analysis.

   \item Computing elastic properties\cite{DCAJ15}, investigating magnetic ordering \cite{HMLP19}, magnetic anisotropy energy (MAE), substitutions, and charge calculations.

   \item {Performing calculations by distorting the structure according to the phonon eigenmodes, examining systems under hydrostatic pressure, and employing various electronic smearing techniques.}

   \item Generating input files for Wannier90\cite{wannier90}, EPW (anisotropic superconductivity)\cite{epw1,epw2}, and WannierTools\cite{wanniertools} calculations, with energies windows provided by users for wannierization.

   \item { Extracting results, including relaxed structures in .cif format, suitable for machine-learning studies.} 
   

\end{itemize}

At present, It offers functionalities like monitoring calculations, aborting jobs, and printing command histories, accessible through straightforward commands. { In this paper, we provide an overview to the package, outline the installation process, and demonstrate its capabilities with examples. More detailed documentation, which includes complete descriptions of inputs, commands, and examples, is included with the package's online documentation [https://neraaz.github.io/HTESP/].}

\section{Methods and Materials}
\begin{figure}[h!]
    \centering
    \includegraphics[scale=0.4]{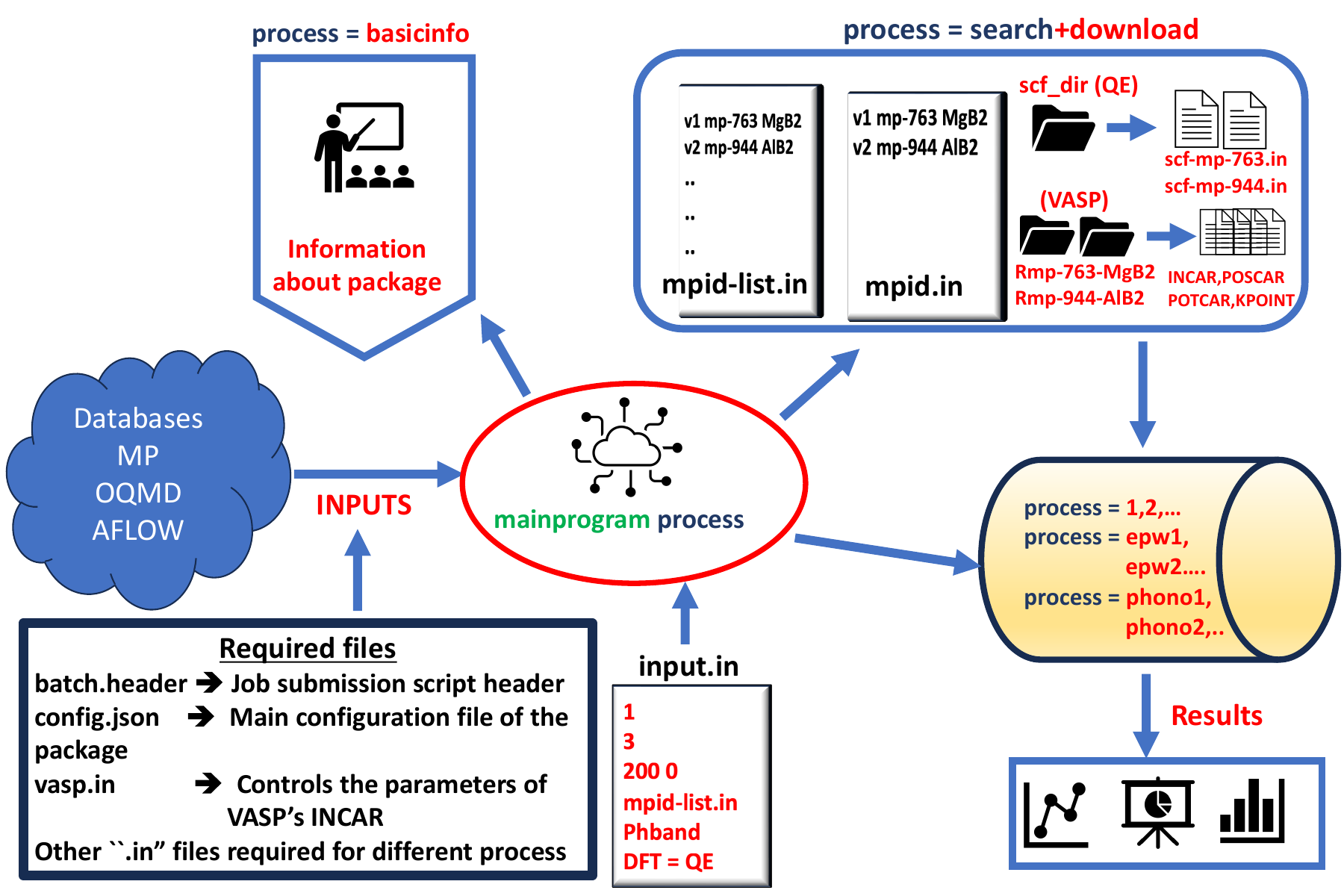}
    \caption{HTESP Package Workflow: Data retrieval from the database is facilitated by a configuration JSON file, config.json, using command-line arguments for search and download operations. Once input files are prepared, users can generate job submission scripts and initiate various processes using the ``mainprogram process" command. Additionally, users can retrieve information about these processes using the ``mainprogram basicinfo" command.}
    \label{fig:overview}
\end{figure}

Figure \ref{fig:overview} illustrates a procedural workflow employed within the package. A complete installation process of the package is presented in Appendix A. The core command at the heart of the package is ``mainprogram", and it can be run with the additional ``process" parameter specified as ``mainprogram process." This ``process" parameter can take the form of numerical values, texts, or a combination of both, each representing a distinct operation. To gain a comprehensive understanding of these processes, users can refer to the ``mainprogram basicinfo" command and follow the instructions provided within it. The workflow starts with extracting materials information from the database using a central input file named ``config.json", which serves as the core input file for the package [Appendix B]. 

The package includes a default version of config.json file with predefined values. This file contains essential details about the materials to be searched within the database, parameters required for preparing input files such as magnetism, convergence parameters, optimization algorthms, etc, pseudopotential information, and other relevant data. For QE, the required input parameters are stored within the ``pwscf\_in" subdictionary. As for VASP, alongside ``config.json" users have the alternative of employing the ``vasp.in" input file to modify the INCAR file acquired from the MPRelaxSet class in pymatgen. Using these files, one can conduct searches and generate input files through a combination of the ``search" and ``download" processes. Two files are produced: ``mpid-list.in" through the ``search" process and ``mpid.in" via the ``download" process. The primary difference between these two files is that ``mpid.in" exclusively contains information about compounds for which input has been generated, constituting a subset of the ``mpid-list.in" file. The generated input files for QE are stored as ``scf-{mpid}.in" within the folder named ``scf\_dir". For VASP, distinct folders named ``R{mpid}-{name}" containing VASP input files are created. Here, ``mpid" refers to the materials IDs, while ``name" denotes the compound's name, which can be found in ``mpid.in" or ``mpid-list.in" files. Furthermore, a ``download.csv" file is generated within the ``download" folder. The columns in this CSV file match the properties listed under the prop keyword [Appendix E]. You can use the MlProcess class in the ml\_processing.py script [from ml\_processing import MlProcess] to easily prepare basic input files for machine learning studies by writing some python codes. Currently, one can use the class to create composition-based features with Matminer \cite{matminer}, generating an `id\_prop.csv` file that includes materials ID and properties necessary for models like Crystal Graph Convolutional Neural Networks (CGCNN) \cite{XG_CGCNN18} and the Atomistic Line Graph Neural Network (ALIGNN) \cite{CD21_ALIGNN}. Although the code currently lacks a command-line interface for this task, we plan to add this feature in the near future with more functionalities. The ``input.in" file enables the execution of various processes for different tasks [Figure~\ref{fig:input_1}.]\\

The ``input.in" file format includes the starting index on the first line, the last index (excluded from calculations) on the second line, the total number of k-points (200) and cutoff index for high-symmetry path (from the last) on the third line, the tracking file ``mpid-list.in" for sequential indices, the plot type denoted by ``phband" (phonon band), and the type of calculations: ``QE" or ``VASP". A batch header file, `batch.header` [Fig.~\ref{fig:input_1}] containing the necessary batch commands related to the computer cluster is required to generate job submission scripts. Comprehensive descriptions of input files and commands are available in the manual or online documentation of the package.

\subsection{Illustrative Examples}
{
In this section, we highlight the capabilities of the HTESP package by performing calculations with QE and VASP code. First, we present the major command arguments to be used with the `mainprogram` script. Next, we demonstrate ground-state electronic structure, phonon properties  and EPC calculations with QE using DFPT for MgB$_2$. We then provide an example of extracting data, preparing input files, and filtering distinct spacegroups for binary Y-C systems. Finally, we showcase command-line automated calculations to compute the convex hull of Y-C systems using VASP code.}

\section{Results}
\subsection{Electron-phonon calculations utilizing QE code}
In this section, we provide an example of computing the phonon-mediated SC properties of MgB$_2$ using the isotropic Eliashberg approximation\cite{E60} and the McMillan-Allen-Dynes formula\cite{MW68,A72,AD72}, as implemented in QE code. One can perform high-throughput calculations for a large number of systems using the ``mpid.in" file. Detailed theory of computing phonon-mediated SC properties is presented in Appendix F. A basic ``config.json" file for initiating QE calculations from a structure file in ``.cif" format appears as the JSON file is shown in Appendix D.\\

Place config.json, batch.header [Fig.~\ref{fig:input_1}], and mp-763.cif (material ID for MgB$_2$ in MP) files inside working directory (workdir). {The pseudopotentials\cite{SSSP18,GBRV14} are stored in the ``pp" folder in the ``{element}.upf" format, where element = Mg or B.} To generate the job submission script (\textbf{mainprogram jobscript}), one can utilize the ``batch.header" file containing necessary batch commands. With this process, the actual commands are appended to this file based on the ``command\_list" key. Some ``run-{command}.sh" files are created, for example, run-scf.sh, run-elph.sh, and so on. Now, generate input file from the ``mp-763.cif" file containing the structure for MgB$_2$ by executing the command ``\textbf{mainprogram download}". This creates scf-mp-763.in file inside the scf\_dir directory. ``input.in" and ``mpid.in" files are also created. Replace ``mpid-list.in" with ``mpid.in" in ``input.in" as;\\

\begin{figure}[h!]
    \centering
    \includegraphics[scale=1.0]{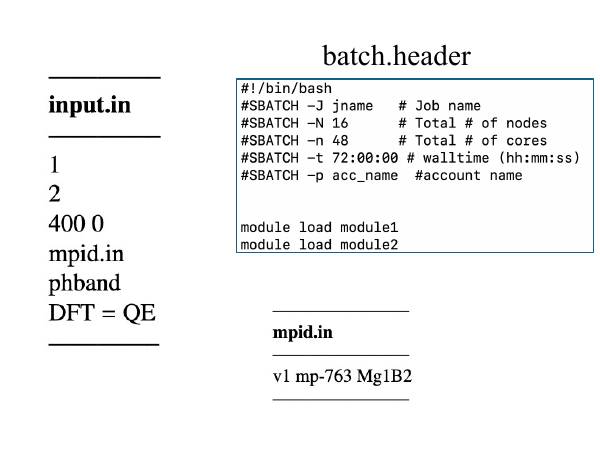}
    \caption{Input control files for QE calculations}
    \label{fig:input_1}
\end{figure}
In this context, ``v1" serves solely as an identifier, ``mp-763" represents the materials ID, and ``Mg1B2" is the name of the compound. With this step, we are ready to perform structure relaxation and further processes. {Throughout the examples in this paper, we approximate the exchange-correlation energy using the generalized gradient approximation (GGA) by Perdew-Burke-Ernzerhof (PBE)\cite{PBE96}.} \\
\begin{figure}[h!]
    \centering
    \includegraphics[scale=0.65]{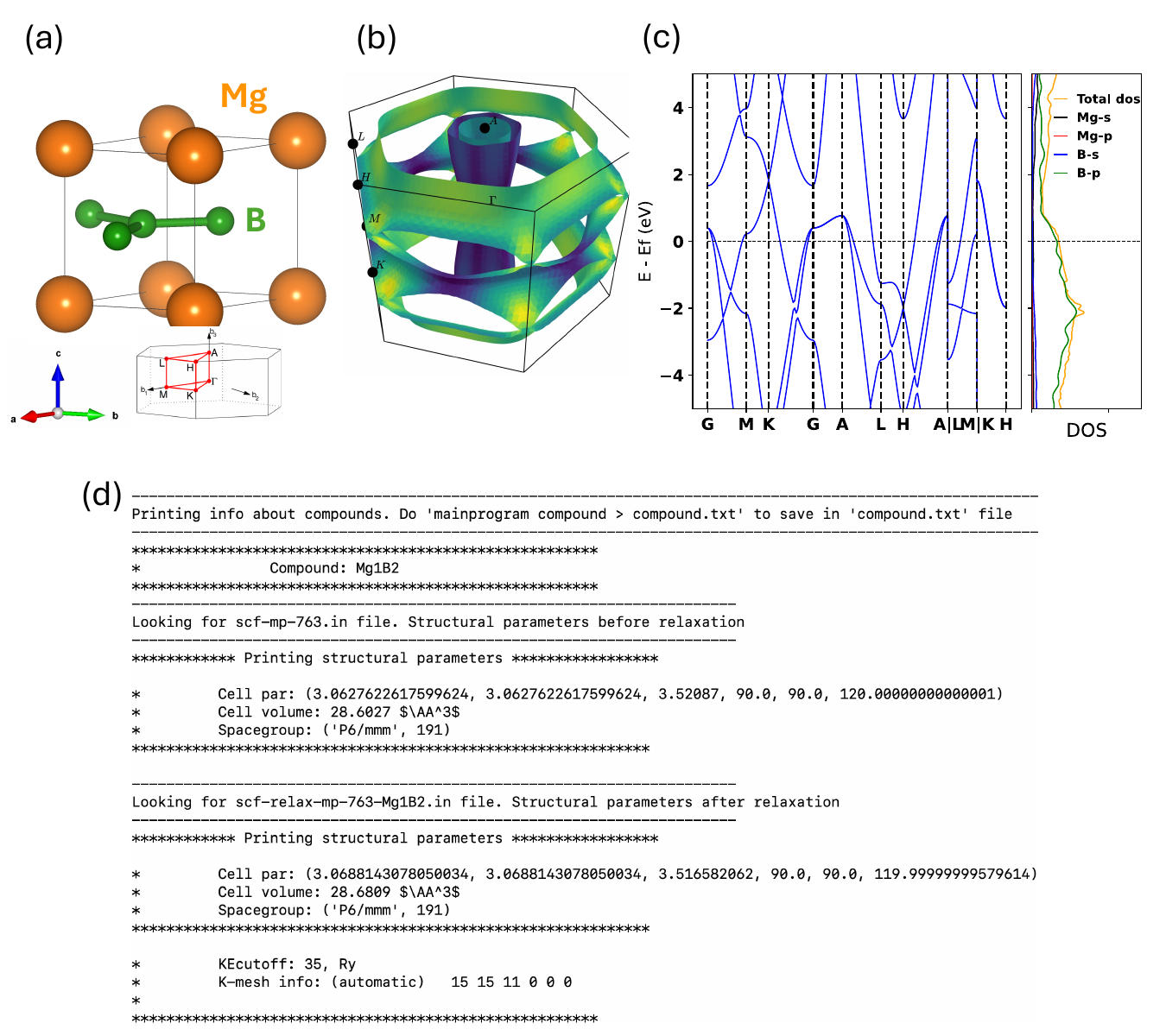}
    \caption{(a) Crystal structure of MgB$_2$ and the first Brillouin-zone showing high-symmetry path (b) 3D Fermi surface (c) Electronic structures (d) Compound information before and after relaxation}
    \label{fig:2}
\end{figure}

\begin{figure}[h!]
    \centering
    \includegraphics[scale=0.8]{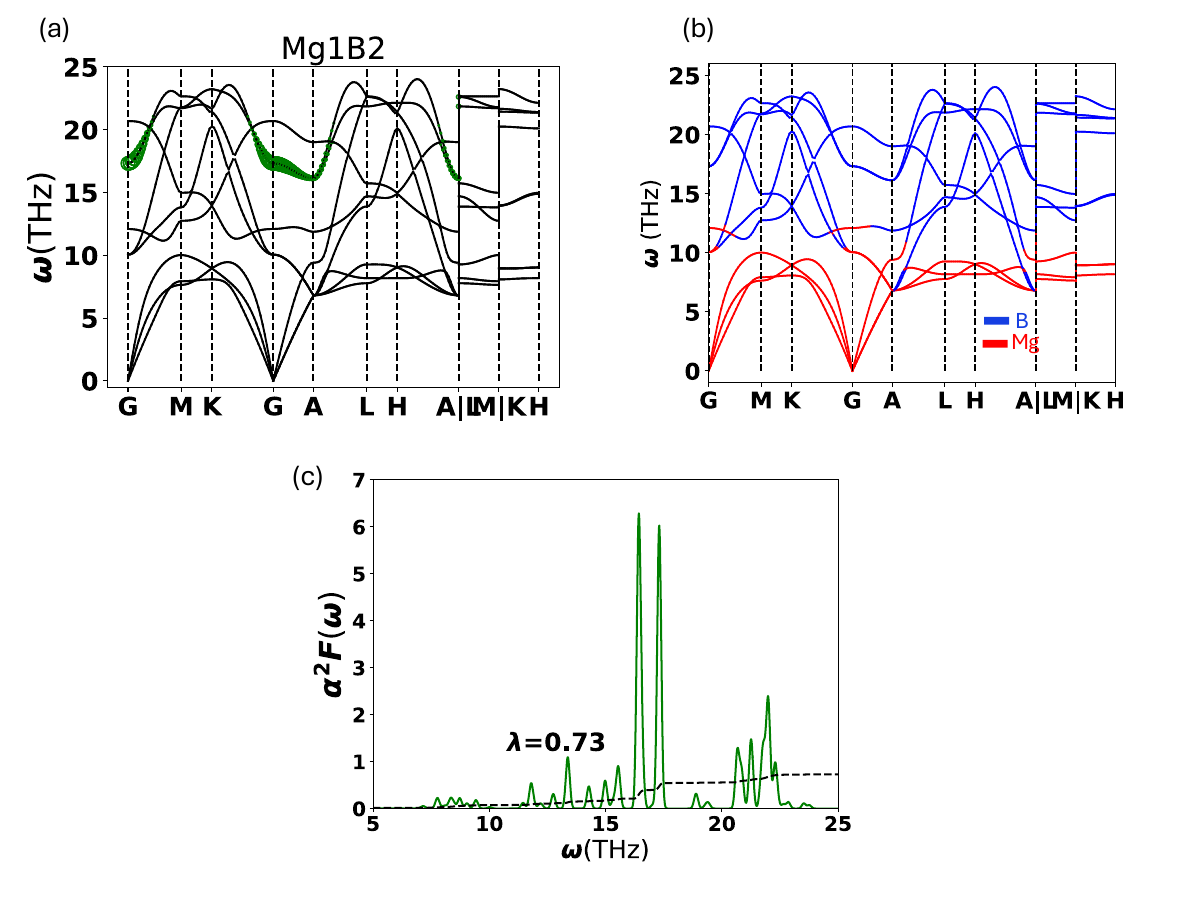}
    \caption{ (a) Phonon-dispersion with mode resolved EPC [see Eqn.(1)] (b) atom-projected phonon-dispersion. Projection is depicted for atomic contributions exceeding 60\%. (c) Eliashberg's spectral function [$\alpha^2$F($\omega$)]}
    \label{fig:3}
\end{figure}
We provide a detailed step-by-step tutorial for computing EPC properties in Appendix G, starting from structure relaxation.
In Figure~\ref{fig:2}, we present structural and electronic properties computed from our command-line automated calculations. The crystal structure has been plotted with VESTA\cite{vesta}, while the BZ plot is from the AFLOW\cite{SC10} web interface (see Figure~\ref{fig:2}(a)). Fermi surface has been plotted with IFermi package utilizing VASP output file ``vasprun.xml" (see Figure~\ref{fig:2}(b)). In Figure~\ref{fig:2}(c) (left panel), we show the electronic band structure of MgB$_2$, demonstrating the metallic behavior (also shown by the DOS/pDOS plot in Figure~\ref{fig:2}(c) (right panel)) and a relatively flat band near the Fermi level ($E_F$) along the $\Gamma$-A direction, consistent with the findings in the Materials Project database\cite{JOHCRD13_MP}. Bands near $E_F$ are dominated by B's 2p orbital. {After each relaxation, the basic structural information about the materials can be checked using the ``mainprogram compound" command. An example of the output is shown in Figure~\ref{fig:2}(d).} Eliashberg's spectral function (Figure~\ref{fig:3}(c)) shows that the largest contribution of $\lambda$ is coming from the 15-20 THz phonon frequency range, which is also confirmed by the mode-resolved $\lambda$-projected phonon dispersion (Figure~\ref{fig:3}(a)) with regions of denser green shades. Additionally, the integrated $\lambda$ for Gaussian smearing of 0.01 Ry is close to 0.73, agreeing with earlier theoretical calculations\cite{SMVG16,SCDFT07}. Similarly, we calculate $\omega_{log}$ to be 762 K, and the critical temperature of 16.8 K using the McMillan-Allen-Dynes formula.

Finally, to analyze the contribution of different atomic species to phonon eigenmodes, we have summed the atomic displacements of each eigenmode for different species 
($|\mathbf{e}^i_{\nu \mathbf{q}}|^2$) as shown in Equation~(\ref{eq-a})\cite{NCW24}, where $\mathbf{e}^{ij}_{\nu \mathbf{q}}$ are the displacements obtained for atom type $i$ from the eigenvector of mode $\nu$ for the $q$-point $\mathbf{q}$ by diagonalizing the dynamical matrix, and the index $j$ ranges from 1 to $N_i$, representing the number of atoms of type $i$. 
\begin{equation}
    |\textbf{e}^i_{\nu \textbf{q}}|^2 = \sum_{j=1}^{N_i}  |\textbf{e}^{ij}_{\nu \textbf{q}}|^2,
    \label{eq-a}
\end{equation}

We color-code the contribution of B in blue, and the contribution from Mg in red. It is evident that vibrations of B contribute significantly to the superconductivity in MgB$_2$ compared to that of Mg. { In addition to ``serial" EPC calculations, you can distribute the calculations in parallel over different \textbf{q} points or \textbf{q}-irr combinations, where ``irr" refers to the index of different irreducible representations of phonon modes. This behavior is controlled by the ``elph\_mode" keyword. The available options are ``serial" (default), ``parallel\_q", and ``parallel\_irr". To use ``parallel\_irr", one should first perform calculations with the ``only\_init" option.}\\

{
\subsection{Combining queries from different database}
\begin{table}[h!]
    \centering
    \caption{Process for extracting data and preparing input files from different database using ``mainprogram process"}
    \begin{tabular}{c|c|c}
    \hline
    \hline
    &  \multicolumn{2}{|c}{process} \\
    \hline
    Database & Search & Download \\
    \hline 
    MP & search & download \\
    OQMD & oqmd-search & oqmd-download \\
    AFLOW & aflow-search & aflow-download \\
    Combine &     & data-combine \\
    \hline 
    \hline 
    \end{tabular}
    \label{tab:1}
\end{table}

\begin{figure}[h!]
    \centering
    \includegraphics[scale=0.6]{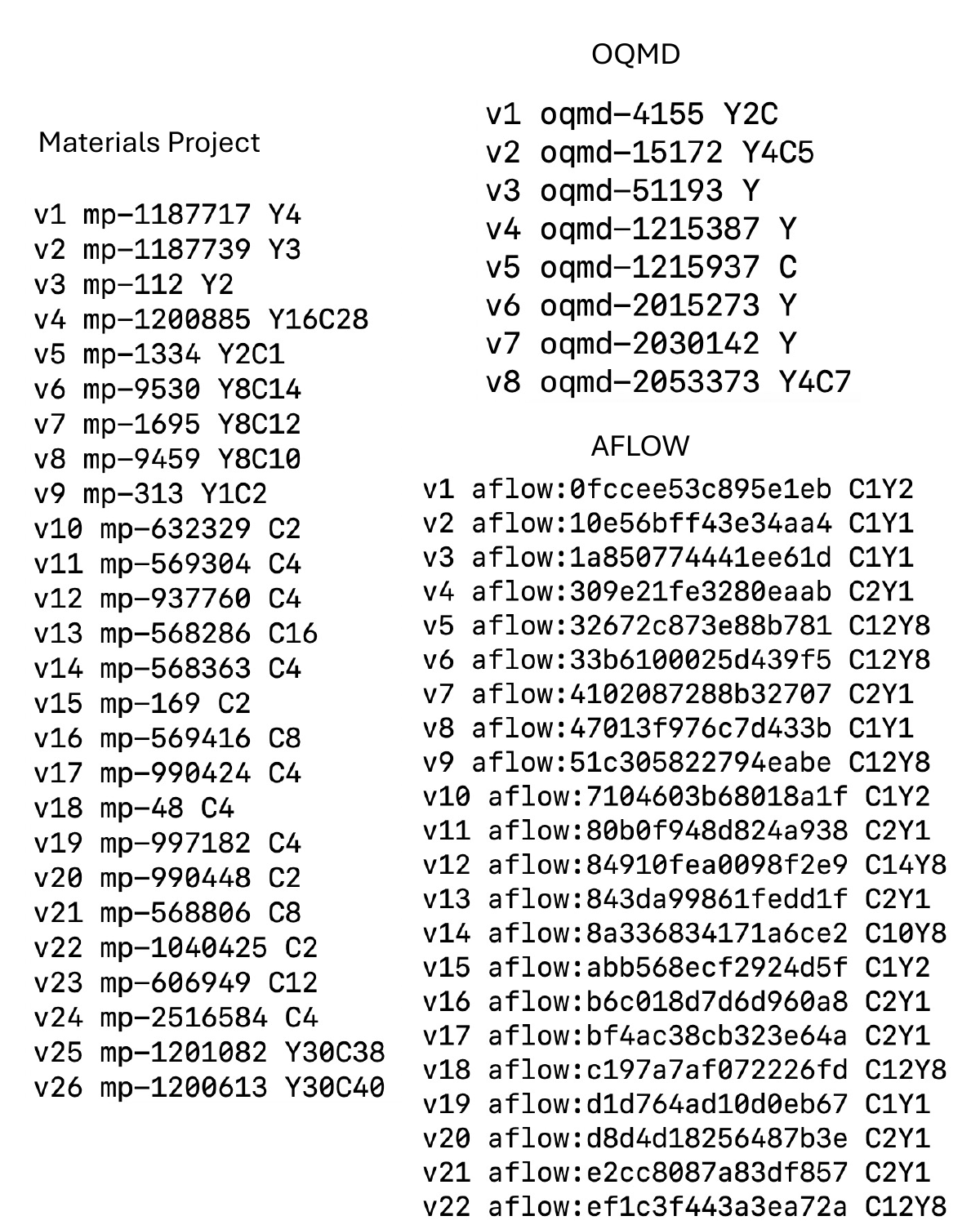}
    \caption{The ``mpid.in" files, obtained after extracting data from 3 different database.}
    \label{fig:download}
\end{figure}

\begin{figure}[h!]
    \centering
    \includegraphics[scale=0.9]{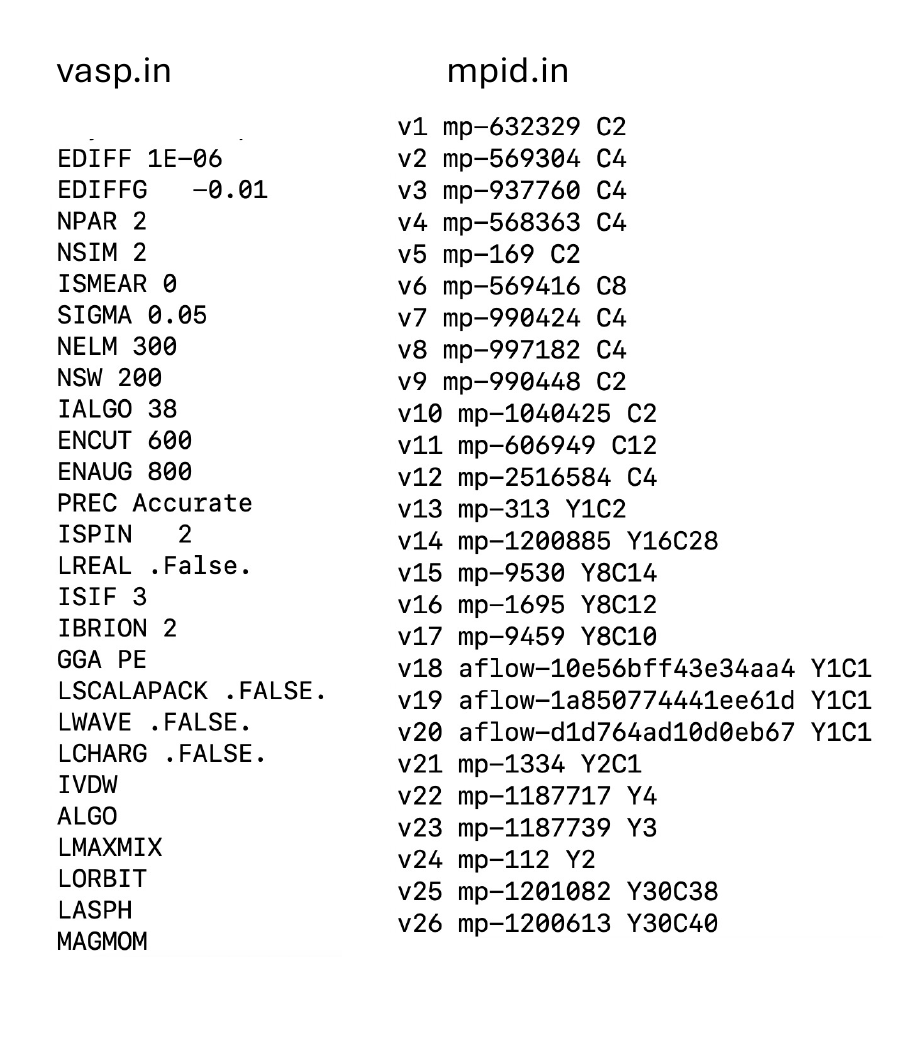}
    \caption{The ``vasp.in" file is required to update INCARs for VASP calculations (left panel). The ``mpid.in" file, obtained after combining data from different databases using ``mainprogram data-combine" command, is shown in the right panel.}
    \label{fig:combine}
\end{figure}
Next, we show the calculations with VASP from preparing input files to submitting calculations, using config.json file as shown in Appendix E.
We outline the procedure for setting up VASP calculations, involving the extraction of data for various compositions of yttrium-carbon (Y-C) from three different databases, combining the data, and preparing systems with distinct spacegroups for VASP computations. Specifically, we extract compounds with ``Y-C" combinations in chemsys modes (``Y", ``C", and ``Y-C" systems) from the MP, OQMD, and AFLOW databases. We filter the MP database to include all combinations with an energy above the ground-state hull by less than 0.08 eV/atom, thereby encompassing a wider range of Y-C compounds. A k-point mesh density of 0.025 is employed. To generate input files, particularly the INCAR, we utilize the ``vasp.in" file [Figure ~\ref{fig:combine} (left panel)] to update the INCAR using the MPRelaxSet class of pymatgen. We incorporate keywords with values to update the INCAR, while removing keys without values from the INCAR. In Table~\ref{tab:1}, we show the commands to search, extract, and prepare input files from these databases, and combine those data according to composition and spacegroup. Specify ``VASP" or ``vasp" for VASP input files when setting up the ``calc" flag. These commands prepare and store INCAR, POSCAR, POTCAR, and KPOINTS in R{mpid}-{name}/relax/ folders for VASP. The materials IDs and compound names are printed in the ``mpid.in" files, as shown in Figure~\ref{fig:download}. Now, merge this data by grouping the structures according to composition and spacegroup to generate a new set of folders and files. The updated ``mpid.in" file is generated [Figure~\ref{fig:combine}(right panel)] along with input files and folders inside the ``filtered\_inputs" folder. In the updated ``mpid.in" file, the OQMD entries have been substituted because of their identical spacegroup found in another database. Subsequent examples will focus exclusively on VASP calculations.

\subsection{Vasp calculations on Y-C systems for thermodynamic phase stability}

\begin{figure}[htbp!]
    \centering
    \includegraphics[scale=0.32]{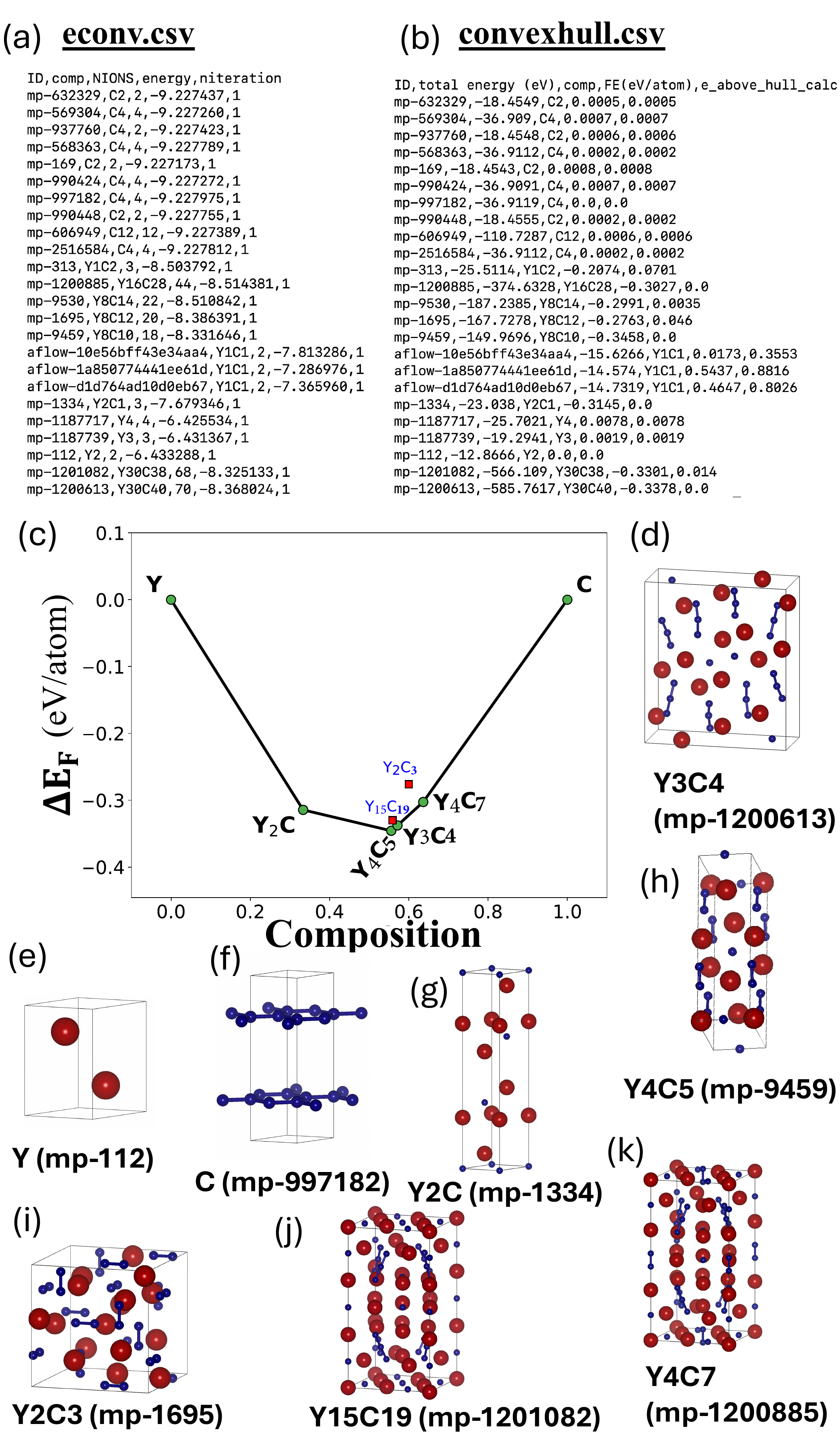}
    \caption{Phase stability of Y-C systems: (a) `econv.csv` file extracted with `\textbf{mainprogram e0}` command. (b) `convexhull.csv` file extracted with `\textbf{mainprogram pd}` command. (c) Convexhull plot with formation energy on y-axis and composition on x-axis. (d)-(k) Various structures obtained from relaxed `.cif` files of the Y-C systems. Only compounds close to convexhull are shown.}
    \label{fig:YC}
\end{figure}
Once we have the input files located in the `R\{mpid\}-\{name\}/relax` folder, which contains the VASP input files, and the corresponding `mpid` and `name` stored in the `mpid.in` file, we can begin the structural relaxation process. Using the `input.in` file with the setting `calc = "VASP"`, we initiate the relaxation procedure following steps 1 through 3, similar to how it is done in QE. Continue repeating process 2 (updating POSCAR) and process 3 (resubmitting) until the code stops further calculations, indicating that the systems have fully relaxed within a single ionic relaxation. This process addresses situations where a calculation is either terminated during optimization or relaxed to a local minimum. Additionally, the input parameters can be continuously adjusted to handle different types of errors during the optimization process. The relaxed structure can be stored in CIF format inside the `cif` folder, and the total energies per ion and the number of ionic iterations can be recorded in `econv.csv` using the `\textbf{mainprogram e0}` command, as shown in Figure~\ref{fig:YC}(a). By executing `\textbf{mainprogram pd}`, we can compute convexhull phase diagram of Y-C systems via pymatgen. The command produces `convexhull.csv` [Figure~\ref{fig:YC}(b)] file and `convexhull.pdf` plot.
The convex hull representing the isothermal phase diagram is crucial for determining the stability of compounds at ambient pressure and temperature.

}
\section{Conclusions}
In summary, we present the HTESP package, which streamlines data extraction from multiple database and automates \textit{ab initio} calculations via a command-line interface. We detail the installation process and showcase the package's high-throughput capabilities through electron-phonon coupling calculations for MgB$_2$ and thermodynamic phase stability convex hull for Y-C system. In addition to its capabilities in phonon-mediated superconductivity and ground state convex hull phase diagram calculations, the package also offers tutorials embedded within the code showing its various other functionalities. This highlights the potential of the HTESP package as a valuable toolkit for researchers, particularly those work on materials exploration with high-throughput screening. { In the future, we plan to enhance the code by adding more functionalities, particularly in plotting, expanding the package to handle 2D materials, and integrating the package with widely used machine learning models such as CGCNN\cite{XG_CGCNN18} and ALIGNN\cite{CD21_ALIGNN}.} Through a combination of bash and Python scripts, our package provides valuable assistance to newcomers in navigating different databases and conducting high-throughput \textit{ab initio} calculations. In addition to the software package, comprehensive examples and online documentation are available to facilitate the learning process (https://github.com/Neraaz/HTESP).

\section{Declaration of competing interest}
The authors declare no competing interests.
\section{Acknowledgements}
We acknowledge the helpful discussion with Dr. James R. Morris at Ames National Laboratory. This work was supported by Ames National Laboratory LDRD and U.S. Department of Energy, Office of Basic Energy Science, Division of Materials Sciences and Engineering. Ames National Laboratory is operated for the U.S. Department of Energy by Iowa State University under Contract No. DE-AC02-07CH11358.

\section{Data availability}
The raw data required to reproduce these findings are available to download from \url{https://github.com/Neraaz/HTESP/tree/main/examples}. The processed data required to reproduce these findings are available to download from \url{https://github.com/Neraaz/HTESP/tree/main/examples}.


\appendix

\section{Installation}
\noindent In this section, we outline the installation procedure for the package.

\subsection{Download Software}

\begin{verbatim}
git clone https://github.com/Neraaz/HTESP.git
\end{verbatim}

Go to HTESP directory,

\begin{verbatim}
cd HTESP
\end{verbatim}

\subsection{Conda Environment}
\begin{verbatim}
conda create --name myenv python==3.9.12
source activate myenv
\end{verbatim}
\subsection{Install dependencies}
\begin{verbatim}
pip install -r requirements.txt
\end{verbatim}

\noindent Install Phonopy (optional) to perform Phonopy calculations.

\subsection{Install HTESP Package}
\begin{verbatim}
pip install . (pip install dot)
\end{verbatim}

\subsection{Check Executable}

\begin{verbatim}
which mainprogram
\end{verbatim}

\noindent Execute ``mainprogram basicinfo" to begin [Appendix C].

\noindent Alternatively, install the development version using the following command:

\begin{verbatim}
python setup.py develop
\end{verbatim}

\begin{verbatim}
which mainprogram
\end{verbatim}

\subsection{After Installation}

\begin{verbatim}
Provide the path to HTESP/src/ and HTESP/src/bash folder in ~/.bashrc
export PATH="path_to_HTESP/src/bash:$PATH"

Provide the path to src file
export PYTHONPATH="path_to_HTESP/src:$PYTHONPATH"
\end{verbatim}

\section{Skeleton of the config.json file}
\begin{lstlisting}[language=json,caption={}]
    {
  "job_script": {},
  "mpi_key": {},
  "download": {,
    "mode": "element"
    "element": {},
    "inp": {},
    "chemsys": {},
    "oqmd": {},
    "aflow": {},
  },
  "conv_test": {},
  "magmom": {},
  "pseudo": {
    "pot": {},
    "PSEUDO": {}
  },
  "substitute": {},
  "pwscf_in": {},
  "strain": [],
  "wanniertools_input": {},
 "kptden":0.025,
 "elph_mode": "serial",
 "plot": {}
}

\end{lstlisting}

\section{Printing information about package}
Once installed, you can execute ``mainprogram basicinfo" to obtain necessary information about the package functionality. Here, the package shows the different processes and their corresponding operations. These key processes encompass further subprocesses, which can be explored by diving deeper into the functionalities of the package.
\begin{lstlisting}[style=bashstyle]

# run 'mainprogram process'

#Look for config.json file in utility/input_files/ and copy that to the working directory

process = jobscript generates the job scripts for the calculations

process = search, search for data in materials project database

process = download, download QE and VASP input files

process = oqmd-search, search for data in oqmd database

process = oqmd-download, download QE and VASP input files

process = aflow-search, search for data in aflow database

process = aflow-download, download QE and VASP input files

process = data-combine, combining and eliminating duplicate inputs for different database

process = process-info, for information about QE+VASP calculations

process = epw-info for EPW input files and wannier90 input files preparation

process = wt-info for wanniertools input files preparation  

process = elastic-input, to create input files with deformations

process = compute-elastic, to compute elastic properties

process = magenum, to create input files for different magnetic state

process = magmom_extract, extract magnetic moment

process = fermisurface to plot fermi surface from vasprun.xml

process = charge-input for creating input files for system with non-zero net charge

process = pressure-input for creating input files for different pressure
#For pressure, 'pressure.in' file is provided with v1 pressure1, v2 pressure2 on different lines. One can also scale volume using scaling factor for isotropic volume change  with v1 scale1, v2 scale 2, on different lines, where scale can be 0.94, 0.96, .. etc. A mpid-pressure.in file is created and pressure value is inserted to scf\_dir/scf-mpid.in files to get scf\_dir/scf-mpid-pressure.in

process = compound, to print information about compound after relaxation

process = checkph, to check the status of el-ph calculation

process = checkfreq, to check negative frequency in the phonon band

process = singlemode provides info for single-mode phonon calculations.
#Also do 'mainprogram process-info' and look for process 23-25 for automated calculations. It requires run-dynmat.sh and run-scf.sh files in working directory

#For POTCAR. Suppose we have POTCARS as  POT_GGA_PAW_PBE/Mg_p/POTCAR
                          #pmg config -p /path_to/POT_GGA_PAW_PBE PBE52
                          # After that add path to .pmgrc.yaml
                          #pmg config --add PMG_VASP_PSP_DIR PBE52

process = change\_k, to change kmesh. Use kpoint.in similar to qpoint.in file. Either provide new kmesh or fractional number to scale old k-mesh

process = history to print latest 10 mainprogram command executed

#First execute 'history -a' in command line before process = history. For MacOS, replace it by ~/.zsh_history in the mainprogram file. 

#Please refer to the Online Documentation for further information and guidance.


\end{lstlisting}

\section{A config.json file for QE calculations}
\begin{lstlisting}[language=json,caption={}]
{
  "job_script": {
  "batch":"batch.header",
  "which_calc": "qe",
  "parallel_command": "mpirun",
  "nproc": "24",
  "command_list":["scf","elph","q2r","matdyn","matdyn-dos","lambda"],
  "command_combine":false,
  "calc_visible_with":"id"
  },
  "mpi_key": {
    "API_KEY": {
      "key": "Use your materials project key"
    }
  },
  "download": {
   "mode": "fromcif",
    "inp": {
      "start": 1,
      "end": 65,
      "nkpt": 200,
      "evenkpt": true,
      "plot": "phband",
      "calc": "QE"
    }
    },
  "pseudo": {
    "PSEUDO": {
      "Mg": 30,
      "B": 35
    }
  },
  "pwscf_in": {
    "magnetic": false,
    "control": {
      "calculation": "vc-relax",
      "nstep": 300,
      "restart_mode": "from_scratch",
      "pseudo_dir": "../../pp/",
      "outdir": "./",
      "tprnfor": true,
      "tstress": true,
      "etot_conv_thr": 1e-05,
      "forc_conv_thr": 0.0001
    },
    "system": {
      "smearing": "gauss",
      "occupations": "smearing",
      "degauss": 0.02
    },
    "electrons": {
      "diagonalization": "david",
      "mixing_mode": "plain",
      "mixing_beta": 0.7,
      "conv_thr": 1e-16,
      "electron_maxstep": 300
    }
  },
 "kptden":0.025,
 "elph_mode": "serial",
 "plot": {
   "atomproj": 0.6,
   "xlim": [-8,10],
   "ylim": [-20,20]
  }
}

\end{lstlisting}

\begin{itemize}
    \item \textbf{job\_script:} It defines the method for creating job submission scripts for execution on the computer cluster.
    \item \textbf{mpi\_key:} It stores the MP API key.
    \item \textbf{download:} It determines the generation of input files and ``input.in".
    \item \textbf{pseudo:} It specifies the kinetic energy cutoff for elements within materials in Ry. ``PSEUDO" key is used for QE, while "pot" key is used for VASP.
    \item \textbf{pwscf\_in:} It contains parameters for QE input files.
    \item \textbf{kptden:} It represents the k-point density.
    \item \textbf{elph\_mode:} Mode of EPC calculations such as serial or parallel.
    \item \textbf{plot:} It includes some basic parameters for plotting.
\end{itemize}
\section{A config.json file for VASP calculations}
In this section, we hightlight the relevent portion of the config.json file to search and extract data and prepare input files for the queries from different database. 

\begin{lstlisting}[language=json,caption={config.json file}]
    {
      "download": {
      "mode": "chemsys",
        "element": {
           "metal": false,
           "FE": false,
           "thermo_stable": false,
           "exclude": ["Lu"],
           "ntype": [1, 2],
           "elm": ["B"],
           "prop": ["material_id", "formula_pretty", "structure", "formation_energy_per_atom", "band_gap", "energy_above_hull", "total_magnetization", "ordering", "total_magnetization_normalized_formula_units", "num_magnetic_sites", "theoretical", "nsites"],
           "ordering": "NM",
           "nsites": 10,
           "spacegroup": null
        },
       "inp": {
           "start": 1,
           "end": 65,
           "nkpt": 200,
           "evenkpt": false,
           "plot": "phband",
           "calc": "VASP",
           "use_cif2cell": false
       },
       "chemsys": {
           "entries": ["Y", "C"],
           "must_include": ["Y","C"],
           "thermo_stable": 0.08,
           "size_constraint": 80,
           "ntype_constraint": 3,
           "form_en": false,
           "metal": false,
           "magnetic": true,
           "spacegroup": null
        },
       "oqmd": {
           "limit": 5000,
           "entries": ["Y", "C"],
           "size_constraint": 60,
           "ntype_constraint": 3,
           "must_include": [],
           "form_en": true,
           "metal": false,
           "magnetic": true,
           "spacegroup": null,
           "thermo_stable": true,
           "FE": true,
           "prop": ["composition", "spacegroup", "volume", "band_gap", "stability"]
        },
        "aflow": {
           "elm": ["Y","C"],
           "nelm": 2,
           "nsites": 60,
           "metal": false,
           "FE": false,
           "spacegroup": null,
           "filter": false,
           "limit": 5000,
           "prop": [
              "spacegroup_relax", "Pearson_symbol_relax"
            ]
         },
  "conv_test": {
     "param": "ecut",
     "ecut": [400, 500, 600],
     "kpoint": [[6, 6, 6], [12, 12, 12], [18, 18, 18]]
    },
  "magmom": {
    "type": "",
    "magmom": {
      "Y": 1,
      "C": 0
    }
    },
  "pseudo": {
    "pot": {
      "C":"C",
      "Y":"Y_sv"
     }
   }, 
   "kptden": 0.025,
   "chull_cutoff": 0.05}

\end{lstlisting}
\begin{itemize}
    \item \textbf{mode: } Mode of the input generation
    \item \textbf{element: } Elemental mode of MP database
    \item \textbf{inp: } Parameters to create ``input.in" file
    \item \textbf{chemsys: } Chemsys mode within MP database
    \item \textbf{oqmd: } Parameters for OQMD database
    \item \textbf{aflow: } Parameters for AFLOW database
    \item \textbf{conv\_test: } Parameters for convergence tests
    \item \textbf{magmom: } Initial magnetic moments in ferromagnetic configuration.
    \item \textbf{pseudo: } Pseudopotential specification
    
\end{itemize}
Here, the term ``mode" specifies how input files are generated, either from the MP or from structure files like ``.cif" or VASP POSCAR files in ``.vasp" format. The ``element" dictionary contains parameters for searching, extracting, and preparing input files from the MP database in ``element" search mode, which searches based on elements or lists of elements. The ``inp" dictionary contains parameters for constructing the ``input.in" file and includes a few other flags. The ``chemsys" mode activates the ``chemsys" dictionary to prepare input files, and conduct a chemical system search in the MP database. For instance, specifying ``Y-C" will search for all elemental and binary compounds within this combination. Lastly, the ``oqmd" and ``aflow" dictionaries hold parameters for searching, extracting, and preparing input files from the OQMD and AFLOW databases, respectively. These dictionaries incorporate various filters such as thermodynamic stability, spacegroup, and metallicity. 

\section{Theory of superconductivity}
Estimating $T_c$ requires $\lambda$ given by
    
\begin{align}
\label{eq1}
    \lambda &=  \sum_{\nu} \int_{BZ} \frac{d\textbf{q}}{\Omega_{BZ}} \lambda_{\textbf{q}\nu},
\end{align}
with mode-resolved $\lambda_{\textbf{q}\nu}$ as,
\begin{align}
    \lambda_{\textbf{q}\nu} &= \frac{1}{N(\epsilon_F)} \sum_{nm}  \int_{BZ} \frac{d\textbf{k}}{\Omega_{BZ}} \frac{|g_{mn,\nu}(\textbf{k},\textbf{q})|^2}{\omega_{\textbf{q}\nu}} \delta(\epsilon_{n\textbf{k}} - \epsilon_F) \delta(\epsilon_{m\textbf{k+q}} - \epsilon_F),
\end{align}

where, $\Omega_{BZ}$ is the volume of the Brillouin zone (BZ), $N(\epsilon_F)$ is the density of states at the Fermi-level ($\epsilon_F$), $g_{mn,\nu}(\textbf{k},\textbf{q})$ is EPC matrix, $\omega_{\textbf{q}\nu}$ is phonon frequency for wavevector \textbf{q} and mode $\nu$, and $\epsilon_{n\textbf{k}}$'s are DFT ground-state eigenvalues for band index n and wavevector \textbf{k}. The double-delta integration around the $\epsilon_F$ over the BZ converges slowly with respect to \textbf{k}- and \textbf{q}-grids, and hence requires denser meshes for converged calculations . To decrease the computational complexity, one can employ the Gaussian broadening techniques \cite{WdG05} by smearing out the discrete states to continuous ones with respect to the $\epsilon_F$.  Eq.(~\ref{eq1}) with Gaussian broadening becomes,
\begin{align}\label{eq2}
    \lambda & \approx \frac{1}{N(\epsilon_F ) N_\textbf{q} N_\textbf{k}} \sum_{nm} \sum_{\nu} \sum_\textbf{q} \sum_\textbf{k} \frac{|g_{mn,\nu}(\textbf{k},\textbf{q})|^2}{\omega_{\textbf{q}\nu}} \frac{1}{2\pi \sigma^2 } exp \left [ -\frac{(\epsilon_{n\textbf{k}} - \epsilon_F)^2 + ( \epsilon_{m\textbf{k}+\textbf{q}} - \epsilon_F)^2}{\sigma^2} \right ].
\end{align}
Here, $N_\textbf{q}$ and $N_\textbf{k}$ respectively are the total number of \textbf{q} and \textbf{k} grid points, and $\sigma$ is the smearing used to broaden states at $\epsilon_F$. With infinitely large \textbf{k}- grids, and $\sigma \rightarrow 0$, the double summation changes back to double delta integration. However, in practice, $|g_{mn,\nu}(\textbf{k},\textbf{q})|^2$ is computed for a reasonable coarse \textbf{k}- and \textbf{q}- grids and then interpolate to fine \textbf{k}- and \textbf{q}- grids to achieve numerical convergence, as implemented in QE\cite{WdG05}. Employing extremely fine grids for interpolation can increase computational cost significantly. Therefore, we have restricted ourselves for choosing fine \textbf{k}-grid only twice of the corresponding coarse grid, which is mostly sufficient \cite{NCW24}.

We also have
\begin{equation}\label{eq3}
      \sum_{mn}  \sum_\textbf{k} \sum_\textbf{q} \delta(\epsilon_{n\textbf{k}} - \epsilon_F) \delta(\epsilon_{m\textbf{k+q}} - \epsilon_F) \sim [N(\epsilon_F)]^2
\end{equation}
resulting, $\lambda$ $\sim$ $N(\epsilon_F)$. Moreover, $\lambda$ can also be obtained from frequency $\omega$ resolved Eliashberg spectral function as \cite{MW68}
\begin{equation}
    \lambda = 2 \int \frac{d\omega \alpha^2F(\omega)}{\omega}
\end{equation}

with the Eliashberg spectral function $\alpha^2F(\omega)$, defined as

\begin{equation}
    \alpha^2F(\omega) = \frac{1}{2} \sum_\nu \int_{BZ} \frac{d\textbf{q}}{\Omega_{BZ}} \omega_{\textbf{q}\nu} \lambda_{\textbf{q}\nu} \delta(\omega - \omega_{\textbf{q}\nu})
\end{equation}

The critical temperature can be calculated using the McMillan-Allen-Dynes formula \cite{AD72},
\begin{equation}\label{Tc}
    T_c = \frac{\omega_{log}}{1.2} exp \left[- \frac{1.04 (1 + \lambda)}{\lambda - \mu^*_c(1 + 0.62\lambda)}  \right],
\end{equation}
where, $\omega_{log} = exp \left[\frac{2}{\lambda} \int \frac{d\omega}{\omega} \alpha^2F(\omega) log\omega\right]$ with frequency $\omega$ resolved Eliashberg spectral function $\alpha^2F(\omega)$, and $\mu^*_c$ is the Coulomb potential, usually in the range of 0.10 to 0.16.

\section{Computing EPC properties of MgB$_2$: step by step tutorial}
\begin{itemize}
    \item \textbf{mainprogram 1:} Initiate structure relaxation
    \item \textbf{mainprogram compound:} Printing structure information before and after the relaxation [Fig.~\ref{fig:2}(d)].
    \item \textbf{mainprogram 2:} Update input with relaxed structure
    \item \textbf{mainprogram 3:} Rerun the relaxation. Repeat processes 2 and 3 until the structure is fully relaxed.
    
    \item \textbf{mainprogram 4:} Prepare inputs required for further processes
    \item \textbf{mainprogram 5:} Perform self-consistent field (SCF) calculations with a fine k-mesh grid
    \item \textbf{mainprogram 6:} Perform SCF calculations with a coarser k-mesh grid
    \item \textbf{mainprogram 7:} Submit electron-phonon coupling (EPC) calculation
    \item \textbf{mainprogram 8:} Process force constants in real space
    \item \textbf{mainprogram 9:} Calculate phonon dispersion in high-symmetry Brillouin-zone (BZ) path
    \item \textbf{mainprogram 10:} Calculate phonon density of states and other EPC quantities
    \item \textbf{mainprogram 11:} Calculate critical temperature (T$_c$), a logarithmic average of phonon frequencies ($\omega_{log}$), and EPC strength constant ($\lambda$)
    \item \textbf{mainprogram 12:} Process phonon dispersion 
    \item \textbf{mainprogram 13:} scf calculation for band calculation
    \item \textbf{mainprogram 14:} Performing band calculation on high-symmetry BZ path
    \item \textbf{mainprogram 15:} Processing band dispersion
    \item \textbf{mainprogram 16:} Density of states (DOS) calculation
    \item \textbf{mainprogram 17:} DOS processing
    \item \textbf{mainprogram 18:} Partial DOS (pDOS) calculation
    \item \textbf{mainprogram 19:} Obtain plots. Available options: phonon band (phband)[Fig.~\ref{fig:3}(a)], mode-resolved $\lambda$-projected phonon dispersion (gammaband)[Fig.~\ref{fig:3}(a)], atom-projected phonon dispersion (phonproj)[Fig.~\ref{fig:3}(b)], eband (electronic band structure) [Fig.~\ref{fig:2}(c)], pdos (DOS and pDOS) [Fig.~\ref{fig:2}(c)]
    
\end{itemize}










%

\end{document}